\documentclass[conference]{IEEEtran}
\IEEEoverridecommandlockouts

\usepackage{cite}
\usepackage{amsmath,amssymb,amsfonts}
\usepackage{algorithm}
\usepackage[noend]{algpseudocode}
\usepackage{graphicx}
\usepackage{textcomp}
\usepackage{xcolor}
\usepackage{array}
\usepackage{booktabs}
\usepackage{multirow}
\usepackage{comment}
\usepackage{caption}
\usepackage{subcaption}
\usepackage{tabularx}
\usepackage{bm}

\usepackage{tikz}
\usepackage{pgfplots}
\pgfplotsset{scaled y ticks=false}
\usepackage{mathrsfs}
\usepackage{mathtools}
\usetikzlibrary{arrows}
\usepgfplotslibrary{external}
\usepackage{xcolor}
\usepackage{pifont}
\pgfplotsset{grid style={dashed,gray}}
\pgfplotsset{minor grid style={dotted,gray}}
\pgfplotsset{major grid style={dashed,gray}}

\makeatletter
\newcommand{\multiline}[1]{%
  \begin{tabularx}{\dimexpr\linewidth-\ALG@thistlm}[t]{@{}X@{}}
    #1
  \end{tabularx}
}
\makeatother

\newcommand*{\rom}[1]{\expandafter\@slowromancap\romannumeral #1@}
\makeatother
\allowdisplaybreaks

\usepackage{setspace}
\usepackage{mathtools}
\makeatother
\DeclareMathOperator*{\argmax}{argmax} 
\makeatother
\DeclareMathOperator*{\argmin}{argmin} 

\DeclareMathOperator{\EX}{\mathbb{E}}
\usepackage[T1]{fontenc}
\usepackage{float}

\graphicspath{ {Images/} }

\def\BibTeX{{\rm B\kern-.05em{\sc i\kern-.025em b}\kern-.08em
    T\kern-.1667em\lower.7ex\hbox{E}\kern-.125emX}}

\usepackage[a4paper,
            bindingoffset=0.2in,
            left=1.32cm,
            right=1.34cm,
            top=1.92cm,
            bottom=4.4cm,
            footskip=0.25in]{geometry}

\usepackage{titlesec}
\titlespacing{\section}{0pt}{*0.7}{*0.7} 
\titlespacing{\subsection}{0pt}{*0.6}{*0.6} 
\titlespacing{\subsubsection}{0pt}{*0.5}{*0.5} 

\usepackage{fancyhdr}
\fancypagestyle{plain}{%
  \fancyhf{}
  
  \fancyhead[C]{\small
    IEEE ICC 2025 4th Workshop on Synergies of Communication, Localization, and Sensing towards 6G
  }
}
\begin{document}
\title{
Near-Field RIS-Assisted Localization 

Under Mutual Coupling
}
\author{
\IEEEauthorblockN{Alireza Fadakar\IEEEauthorrefmark{1}, Musa Furkan Keskin\IEEEauthorrefmark{2}, Hui Chen\IEEEauthorrefmark{2}, Henk Wymeersch\IEEEauthorrefmark{2}, Andreas F. Molisch\IEEEauthorrefmark{1}}
\IEEEauthorblockA{\IEEEauthorrefmark{1}University of Southern California, Los Angeles, CA, USA
\\\{fadakarg, molisch\}@usc.edu}
\IEEEauthorblockA{\IEEEauthorrefmark{2}Chalmers University of Technology, Gothenburg, Sweden
\\\{furkan, hui.chen, henkw\}@chalmers.se}
}

\maketitle
\thispagestyle{fancy}                     

\begin{abstract}
Reconfigurable intelligent surfaces (RISs) have the potential to significantly enhance the performance of integrated sensing and communication (ISAC) systems, particularly in line-of-sight (LoS) blockage scenarios. 
However, as larger RISs are integrated into ISAC systems, mutual coupling (MC) effects between RIS elements become more pronounced, leading to a substantial degradation in performance, especially for localization applications. 
In this paper, we first conduct a misspecified and standard Cramér-Rao bound analysis to quantify the impact of MC on localization performance, demonstrating severe degradations in accuracy, especially when MC is ignored. 
Building on this, we propose a novel joint user equipment localization and RIS MC parameter estimation (JLMC) method in near-field wireless systems.
Our two-stage MC-aware approach outperforms classical methods that neglect MC, significantly improving localization accuracy and overall system performance. Simulation results validate the effectiveness and advantages of the proposed method in realistic scenarios.
\end{abstract}

\vspace{-0.2cm}
\section{Introduction}
\vspace{-0.15cm}
The emergence of reconfigurable intelligent surfaces (RISs) with large apertures is poised to be a key enabler of high data rates, low latency, and ubiquitous connectivity in 6G communication systems. 
Recent studies highlight the potential of RISs to address the challenges faced by integrated sensing and communication (ISAC) systems \cite{Liu2023ISAC-RIS}. 
This advancement increases the likelihood of near-field (NF) scenarios, highlighting the need for accurate NF models and algorithms to meet the demanding requirements of communication, localization, and sensing in 6G networks \cite{Chen2024NF, Ozturk2024pixel}. 
In particular, when user equipment (UE) is in close proximity to the RIS, wavefront curvature enables direct localization, even in the absence of a line-of-sight (LoS) link between the UE and the base station (BS) \cite{rinchi2022compressive, Ozturk2024pixel, Ozturk2023Phase, Rahal2024RIS}. 

%
Furthermore, as RIS technology evolves and larger arrays are incorporated into ISAC systems, mutual coupling (MC) becomes an increasingly critical factor, significantly impacting the performance of RIS-assisted systems \cite{Zheng2024Mutual}. 
MC originates from the electromagnetic (EM) field interactions between adjacent RIS elements, where the radiation from one element influences the response of its neighboring elements. 
This phenomenon can significantly impair the performance of RIS-assisted systems, specially in applications such as localization and sensing. 
Therefore, mitigating MC effects is crucial to improving the accuracy and efficiency of RIS-assisted ISAC systems.

Several studies have explored the use of RISs for localization in either far-field (FF) or NF scenarios (e.g., see \cite{fadakar2024multi, Chen2024Multi} for FF and \cite{Ozturk2023Phase, Ozturk2024pixel, Dardari2022LOS} for NF and references therein). 
A common limitation in these works is the omission of MC effects at the RIS, which results in overly optimistic performance estimates. 
To address hardware impairments, \cite{Liu2018direction} applies deep learning for direction-of-arrival estimation in uniform linear arrays (ULAs), while \cite{Rivetti2023endtoend} employs an autoencoder for 2D localization with multiple ULA-equipped BSs, considering effects like MC. 
However, these supervised approaches require extensive labeled data and rely on classical linear modeling for MC, which is impractical, especially for RISs, limiting their applicability in complex 3D scenarios and large-array systems. 

Additionally, among studies addressing hardware impairments in RIS-assisted systems, \cite{Ozturk2023Phase} explores NF localization under phase-dependent amplitude variations at individual RIS elements, while \cite{Ozturk2024pixel} investigates the impact of RIS pixel failures. However, these two works ignore MC between RIS elements. 
Moreover, \cite{Bayraktar2024channel} examines channel estimation and localization in the FF under hardware impairments but relies on an unrealistic linear MC model.  
More recently, \cite{Zheng2024Mutual} experimentally analyzes MC effects among RIS elements using an MC-aware communication model based on scattering matrices. 
The study proposes a practical model training approach that estimates MC parameters from a single 3D full-wave simulation of the RIS radiation pattern. Following this, in \cite{zheng2024mcchannel}, a two-stage approach is proposed to address channel parameter estimation and beamforming challenges in active RIS-assisted communication using sparse recovery techniques.


Unlike previous works, this paper is the first to examine the impact of MC effects in NF RIS localization, and it quantifies the performance degradation of classical techniques when these effects are neglected. The main contributions of this paper are summarized as follows:
\vspace{-0.1cm}
\begin{itemize}
\item \textbf{Investigation of RIS-Aided NF Localization with MC Effects:}
We address the challenge of NF RIS-assisted localization in the presence of MC by adopting a practical end-to-end EM communication model  \cite{Gradoni2021Mutual, Pinjun2024Impact}. 
The MC effects alter the RIS phase profiles, complicating the problem of joint localization and MC parameter estimation (JLMC).
\item \textbf{Impact Evaluation of MC on Localization Accuracy:} 
We apply the misspecified Cramér-Rao bound (MCRB) to assess the impact of MC on localization accuracy, comparing scenarios where the UE is aware or unaware of MC effects. 
Our analysis provides key insights into the conditions including power levels and severity of MC effects under which ignoring MC leads to substantial degradation in localization performance.
\item \textbf{Two-step Algorithm Development for Localization and MC parameter Estimation:}
We propose a two-step algorithm to address the JLMC problem. 
In the first stage, the 2D angles of departure (2D-AOD) from the RIS to the UE, the distance between UE and RIS, and the MC scattering values are initially estimated.
%
In the second stage, an alternating optimization algorithm is introduced to jointly refine the 2D-AOD, distance and MC vector which are then used for UE positioning. 
Extensive simulations validate the superior performance of the proposed method.
\end{itemize}

\section{System Model and Problem Formulation}
This section begins by describing the system model and its geometric configuration. 
Next, we present the signal model and formulate the JLMC problem under MC effects.
\vspace{-0.1cm}
\subsection{Geometry and Signal Model}\label{sec:system-model}
\begin{figure}
\centering
\includegraphics[width=0.8\columnwidth]{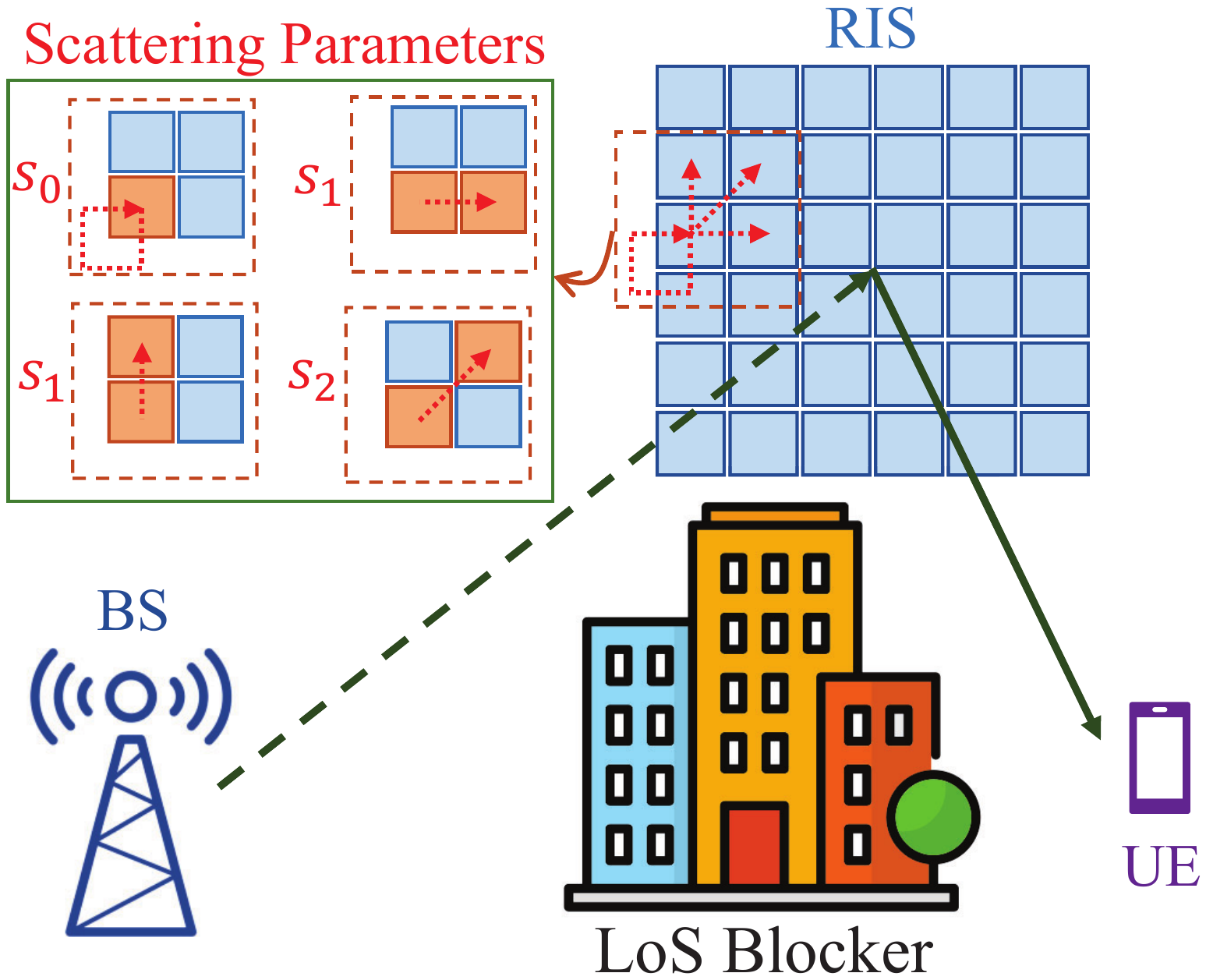}
\caption{
RIS-assisted ISAC system in the near field, where UE localization and MC calibration are performed at the same time. 
The scattering coefficients in \eqref{def:scattering-mat} characterize the MC effects between RIS elements, which are also illustrated.
}
\label{fig:system-model}
\end{figure}

As illustrated in Fig.~\ref{fig:system-model}, consider an ISAC system consisting of a single antenna BS, a passive RIS with $M_r = M_1M_2$ elements, where $M_1$ and $M_2$ denote the number of rows and columns of the RIS, respectively, and a single-antenna UE. 
The positions of the BS, the center of the RIS, and UE are represented by $\bm{p}_b \in \mathbb{R}^3$, $\bm{p}_r \in \mathbb{R}^3$, and $\bm{p}_u \in \mathbb{R}^3$, respectively, while $\bm{p}_{r,n}$ denotes the position of the $n$-th element of the RIS. 
We assume that the LoS path between the BS and UE is blocked, and that the UE is stationary, thereby neglecting Doppler effects for simplicity.

The BS communicates with the UE via the RIS by transmitting $N_t$ narrowband pilot symbols, $\bm{x} = [x_1, \dots, x_{N_t}]^\mathsf{T}$. 
For simplicity, we assume that $x_i = 1$ for all $i$ throughout the paper. 
After stacking the received signals from all $N_t$ transmissions, the downlink received signal vector at the UE, denoted as $\bm{y} \in \mathbb{C}^{N_t}$, is given by
\begin{equation}\label{eq:Y}
\bm{y}
=
\beta
\bm{h}
+
\bm{n},
\end{equation}
where $\bm{n}\sim\mathcal{CN}(0,N_0\bm{I}_{N_t})$ denotes the additive white Gaussian noise vector with $N_0$ being the noise power spectral density (PSD). 
In addition, $\beta$ denotes the overall channel gain which can be modeled as \cite{Ozdogan2020intelligent, Tang2021path, Rahal2024RIS}:
\begin{equation}
\beta
=
\frac{
\lambda^2
\sqrt{PG_tG_r}
\text{exp}(j\varphi)
}{
16\pi^2
\lVert \bm{p}_u-\bm{p}_r\rVert
\lVert \bm{p}_r-\bm{p}_b\rVert
},
\end{equation}
where $P$ represents the transmitted power, $\varphi$ is the global phase offset, $\lambda$ denotes the signal wavelength, and $G_t$ and $G_r$ specify the transmit and receive antenna gains, respectively. 
Furthermore, $\bm{h} \in \mathbb{C}^{N_t}$ is the cascaded RIS-reflected channel vector i.e., accounting for RIS array responses and phase shifts, with its $t$-th element given by
\begin{equation}\label{eq:ch-element}
[\bm{h}]_t
=
\bm{a}(\bm{p}_u)^\mathsf{T}
\bm{\Omega}_t^{'}
\bm{a}(\bm{p}_b),
\end{equation}
For a given position $\bm{p}$, the NF RIS steering vector, denoted as $\bm{a}(\bm{p}) \in \mathbb{C}^{M_r}$, has its $n$-th element defined as
\begin{equation}\label{eq:NF-str-vec}
[\bm{a}(
\bm{p}
)]_n
=
\text{exp}
\left(
-jk
(
\lVert \bm{p}-\bm{p}_{r,n}\rVert
-
\lVert \bm{p}-\bm{p}_{r}\rVert
)
\right),
\end{equation}
where $k=2\pi/\lambda$ is the wave number.
In \eqref{eq:ch-element}, $\bm{\Omega}^{'}_t \in \mathbb{C}^{M_r \times M_r}$ represents the unknown MC-affected phase control coefficients of the RIS during the $t$-th transmission, \cite[Eq.~(5)]{Li2024Beyond}, \cite[Eq.~(8)]{Abrardo2024S_param}, \cite[Eq.~(7)]{Zheng2024Mutual}:
\begin{equation}\label{def:Omega-definition}
\bm{\Omega}_{t}^{'}
=
(
\bm{\Omega}_{t}^{-1}
-
\bm{S}
)^{-1},
\end{equation}
where $\bm{S} \in \mathbb{C}^{M_r \times M_r}$ denotes the scattering matrix. It is crucial to highlight that the matrices $\bm{\Omega}_t$ are controlled by the system designer, and therefore, they are known in advance. 
In contrast, the scattering matrix $\bm{S}$ is unknown a priori, which makes the MC-affected RIS phase profiles $\bm{\Omega}_t^{'}$ also unknown.
According to \cite{Pinjun2024Impact}, RIS elements that are closer together exhibit higher mutual impedance, which results in increased scattering between them. 
Consequently, the scattering matrix $\bm{S}$ can be approximated by assuming negligible mutual impedance for elements that are sufficiently far apart \cite[Eq.~(15)]{Zheng2024Mutual}:
\begin{equation}\label{def:scattering-mat}
\bm{S}
\approx
\sum_{i=0}^{N_m-1}s_{i}\bm{A}_{i}.
\end{equation}
where $N_m$ is the number of the dominant MC scattering parameters $s_0,s_1,\dots ,s_{N_m-1}$. 
In addition, $\bm{A}_{i} \in \{0,1\}^{M_r \times M_r}$ denotes the support matrix for the MC coefficient $s_{i} \in \mathbb{C}$. 
Specifically, $[\bm{A}_{i}]_{k,l} = 1$ if the voltage wave $s_{i}$ is measured at the $k$-th element of the RIS when a unit voltage wave is applied at the $l$-th element of the RIS, and zero otherwise.
\vspace{-0.4cm}
\subsection{Problem Formulation}
\vspace{-0.3cm}
The objective of this paper is to jointly estimate the MC parameter vector $\bm{s}=[s_0,\dots ,s_{N_m-1}]^\mathsf{T}$ and the UE position $\bm{p}_u$. 
The ML estimator for the JLMC problem is given by
\begin{equation}\label{eq:optimization-problem}
(\hat{\beta},\hat{\bm{p}}_u,\hat{\bm{s}})
=
\argmin_{\beta,\bm{p}_u,\bm{s}}
\lVert \bm{y}-\beta\bm{h}\rVert^2.
\end{equation}
The closed form optimal complex value of $\beta$ for a given $\bm{p}_u$ can be easily obtained using least squares as:
\begin{equation}\label{eq:beta-opt}
\hat{\beta}
=
\bm{h}^\mathsf{H}
\bm{y}
/
\lVert
\bm{h}
\rVert^2.
\end{equation}
After inserting \eqref{eq:beta-opt} back into \eqref{eq:optimization-problem} and then removing the terms not dependent on the optimization variables, the problem can be simplified as:
\begin{equation}\label{eq:optimization-problem-simp}
(\hat{\bm{p}}_u,\hat{\bm{s}})
=
\argmax_{\bm{p}_u,\bm{s}}
\lvert\bm{h}^\mathsf{H}\bm{y}_c\rvert^2
/
\lVert
\bm{h}
\rVert^2.
\end{equation}
Before presenting the proposed method, we first quantify the localization performance in the next section using the MCRB and the standard CRB analysis.
\vspace{-0.15cm}
\section{MCRB and CRB Analysis Under MC}
\vspace{-0.1cm}
\subsection{MCRB Analysis}
\vspace{-0.1CM}
In this subsection, we evaluate the localization performance when the UE is unaware of the presence of MC and consequently estimates its location by assuming $\bm{S} = \bm{0}$. 
We employ the MCRB analysis to derive theoretical bounds on the localization accuracy under these conditions.
\vspace{-0.2cm}
\subsubsection{True and Assumed Models}
In the following, we define the true and assumed models. 
The true channel model is expressed as:
\begin{equation}\label{eq:true-model}
\bm{y}
=
\overline{\beta}\ 
\overline{\bm{h}}
+
\bm{n},\ 
[\overline{\bm{h}}]_t
=
\bm{a}(\overline{\bm{p}}_u)^\mathsf{T}
\bm{\Omega}_t^{'}
\bm{a}(\bm{p}_b),
\end{equation}
where $\overline{\beta}$ and $\overline{\bm{p}}_u$ denote the true values of the unknown parameters $\beta$ and the UE position $\bm{p}_u$, respectively. 
For a given MC vector $\bm{s}$, the probability density function (PDF) of the true model in \eqref{eq:true-model} can be expressed as:
{\small
\begin{equation}
p_c(\bm{y})
=
\frac{1}{(\pi N_0)^{N_t}}
\text{exp}
\left(
-\frac{\lVert \bm{y}-\bm{y}_c\rVert^2}{N_0}
\right),
\end{equation}
}
where 
$\bm{y}_c=\overline{\beta}\,
\overline{\bm{h}}
\in \mathbb{C}^{N_t}$.

For the assumed model, we consider an ideal RIS without MC. 
Specifically we assume $\bm{S}=\bm{0}$ which leads to:
\begin{equation}\label{eq:assumed-model}
\bm{y}
=
\beta
\tilde{\bm{h}}
+
\bm{n},\ 
[\tilde{\bm{h}}]_t
=
\bm{a}(\bm{p}_u)^\mathsf{T}
\bm{\Omega}_t
\bm{a}(\bm{p}_b).
\end{equation}
Since $\bm{\Omega}_t$ is diagonal, $\tilde{\bm{h}}$ can be represented in a compact form $\tilde{\bm{h}}^\mathsf{T}=\bm{b}(\bm{p}_u)^\mathsf{T}\bm{W}$, where $\bm{W}\in\mathbb{C}^{M_r\times N_t}$ whose $t$-th column is $[\bm{W}]_{:,t}=\mathrm{diag}(\bm{\Omega}_t)$ and $\bm{b}(\bm{p}_u)=\bm{a}(\bm{p}_u)\odot \bm{a}(\bm{p}_b)$. 
The misspecified PDF is obtained as
{\small
\begin{equation}\label{eq:p_m}
p_m(\bm{y}|\bm{\gamma})
=
\frac{1}{(\pi N_0)^{N_t}}
\text{exp}
\left(
-\frac{\lVert \bm{y}-\bm{y}_m(\bm{\gamma})\rVert^2}{N_0}
\right),
\end{equation}
}
where 
$\bm{\gamma}=[\beta_R,\beta_I,\bm{p}_u^\mathsf{T}]^\mathsf{T}$ is the unknown parameters, where $\beta_R=\Re\{\beta\}$, $\beta_I=\Im\{\beta\}$, and $\bm{y}_m(\bm{\gamma})=\beta
\tilde{\bm{h}}
\in \mathbb{C}^{N_t}$.
The pseudo-true parameter is used in the MCRB derivation and is defined as
{\small
\begin{equation}\label{eq:Pseudo-True}
\bm{\gamma}_0
=
\text{arg}\min_{\bm{\gamma}}
\mathcal{D}(p(\bm{y}) \, \| \, \tilde{p}(\bm{y}|\bm{\gamma})) 
\stackrel{(a)}{=}
\text{arg}\min_{\bm{\gamma}}
\lVert 
\bm{y}_c
-
\bm{y}_m(\bm{\gamma})
\rVert
\end{equation}
}
where (a) is obtained by utilizing \cite[Lemma~1]{Ozturk2023Phase}.

\subsubsection{MCRB and LB}
The lower bound of the covariance matrix for any unbiased estimator of $\bm{\gamma}_0$ with respect to the true vector $\overline{\bm{\gamma}}$ can be obtained as:
\begin{equation}
\EX_{p_c}
\{
(\hat{\bm{\gamma}}(\bm{y}) - \overline{\gamma})
(\hat{\bm{\gamma}}(\bm{y}) - \overline{\gamma})^\mathsf{T}
\}
\succeq
\mathrm{LB}(\bm{\gamma}_0),
\end{equation}
where $\hat{\bm{\gamma}}(\bm{y})$ denotes the unbiased estimator of $\bm{\gamma}_0$ based on \eqref{eq:assumed-model}. 
Moreover, the lower bound matrix is defined as:
\begin{equation}\label{eq:LB_def}
\mathrm{LB}(\bm{\gamma}_0)
=
\underbrace{\bm{A}_{\bm{\gamma}_0}^{-1}
\bm{B}_{\bm{\gamma}_0}
\bm{A}_{\bm{\gamma}_0}^{-1}}_{
\mathrm{MCRB}
(\bm{\gamma}_0)
}
+
\underbrace{
(\overline{\bm{\gamma}}-\bm{\gamma}_0)
(\overline{\bm{\gamma}}-\bm{\gamma}_0)^\mathsf{T}
}_{\mathrm{Bias}(\bm{\gamma}_0)},
\end{equation}
where $\bm{A}_{\bm{\gamma}_0}\in\mathbb{R}^{5\times 5}$ and $\bm{B}_{\bm{\gamma}_0}\in\mathbb{R}^{5\times 5}$ can be calculated as:
{\small
\begin{align}
& 
\bm{A}_{\bm{\gamma}_0}
=
\EX_{p_c}
\bigg\{
\frac{\partial^2}{\partial\bm{\gamma}^2}
\ln
p_{m}(\bm{y}|\bm{\gamma})
\bigg|_{\bm{\gamma}=\bm{\gamma}_0}
\bigg\}, \label{eq:A-def} \\
&
\bm{B}_{\bm{\gamma}_0}
=
\EX_{p_c}
\bigg\{
\frac{\partial \ln p_{m}(\bm{y}|\bm{\gamma})}{\partial \bm{\gamma}}
\left(
\frac{\partial \ln p_{m}(\bm{y}|\bm{\gamma})}{\partial \bm{\gamma}}
\right)^\mathsf{T}
\bigg|_{\bm{\gamma}=\bm{\gamma}_0}
\bigg\} \label{eq:B-def}
\end{align}
}
Thus, the position error bound (PEB) on the localization accuracy under MC when the UE is not aware of MC is determined as:
\begin{equation}\label{eq:PEB_unaware}
\mathrm{PEB}^{\text{unaware}}
=
\mathrm{Tr}\{
[\mathrm{LB}(\bm{\gamma}_0)]_{3:5,3:5}
\}
\end{equation}

\subsubsection{Pseudo-True Parameter Estimation}
In this subsection, we describe how $\bm{\gamma}_0$, $\bm{A}_{\bm{\gamma}_0}$, and $\bm{B}_{\bm{\gamma}_0}$ are determined. 
To compute $\bm{\gamma}_0$, we substitute $\bm{y}_m(\bm{\gamma})$ with its definition from \eqref{eq:assumed-model}, which results in:
\begin{equation}\label{eq:opt-problem}
\bm{\gamma}_0
=
\argmin_{\bm{\gamma}}
\lVert 
\bm{y}_c
-
\beta
\tilde{\bm{h}}(\bm{p}_u)
\rVert^2.
\end{equation}
Using a similar technique which was used to simplify \eqref{eq:optimization-problem}, $[\bm{\gamma}]_{1:2}$ can be determined using a closed form formula similar to \eqref{eq:beta-opt}. 
Thus, the problem transforms to:
\begin{equation}\label{eq:pseudo-opt}
[\bm{\gamma}_0]_{3:5}
=
\argmax_{\bm{p}_u}
\lvert\tilde{\bm{h}}(\bm{p}_u)^\mathsf{H}\bm{y}_c\rvert^2
/
\lVert
\tilde{\bm{h}}(\bm{p}_u)
\rVert^2,
\end{equation}
Note that when the MC effect is sufficiently small (i.e., lower $\lVert \bm{s}\rVert$), the solution of \eqref{eq:pseudo-opt} is close to the true UE position $\overline{\bm{p}}_u$. 
However, for more severe MC, the solution may deviate further from $\overline{\bm{p}}_u$. 
To obtain a more robust solution, we perform a 3D discrete search within a small cube of side length $x_s$ centered at $\overline{\bm{p}}_u$. 
Using the obtained position as the initial estimate, we apply the Nelder-Mead algorithm\cite{FADAKAR2024103382, fadakar2024multi}, an off-the-shelf optimizer, to obtain the gridless, accurate solution.
\subsubsection{MCRB Derivation}
By substituting \eqref{eq:p_m} in \eqref{eq:A-def} and \eqref{eq:B-def}, the first and second order derivatives with respect to $[\bm{\gamma}]_i$ are obtained as:
{\small
\begin{align}
& 
\frac{\partial \ln p_{m}(\bm{y}|\bm{\gamma})}{\partial [\bm{\gamma}]_i}
=
\frac{2}{N_0}
\Re
\{
\dot{\bm{y}}_m(\bm{\gamma})^\mathsf{H}
(\bm{y}-\bm{y}_m(\bm{\gamma}))
\}, \\
&
\begin{aligned}[t]
\frac{\partial^2 \ln p_{m}(\bm{y}|\bm{\gamma})}{\partial [\bm{\gamma}]_i\partial[\bm{\gamma}]_j}
=
\frac{2}{N_0}
\big[
& \Re
\big\{
\dot{\bm{y}}^{'}_m(\bm{\gamma})^\mathsf{H}
(\bm{y}-\bm{y}_m(\bm{\gamma}))
\\
& -
\dot{\bm{y}}_m(\bm{\gamma})^\mathsf{H}
\bm{y}^{'}_m(\bm{\gamma})
\big\}
\big],
\end{aligned}
\end{align}
}
where $\dot{\bm{y}}_m$ and $\bm{y}^{'}_m$ denote the element-wise derivatives of the vector $\bm{y}_m(\bm{\gamma})$ with respect to $[\bm{\gamma}]_i$ and $[\bm{\gamma}]_j$, respectively. 
After performing expectation with respect to the PDF $p_c(\bm{y})$, the elements of the matrices $\bm{A}_{\bm{\gamma}_0}$ and $\bm{B}_{\bm{\gamma}_0}$ can be obtained as:
{\small
\begin{equation*}
\begin{aligned}
&
[\bm{A}_{\bm{\gamma}_0}]_{i,j}
\!\!=\!\!
\frac{2}{N_0}
\Re
\big\{
\dot{\bm{y}}^{'}_m(\bm{\gamma}_0)^\mathsf{H}
(\bm{y}_c-\bm{y}_m(\bm{\gamma}_0))
-
\dot{\bm{y}}_m(\bm{\gamma}_0)^\mathsf{H}
\bm{y}^{'}_m(\bm{\gamma}_0)
\big\},
\\
&
[\bm{B}_{\bm{\gamma}_0}]_{i,j}
\!\!=\!\!
\frac{4}{N_0^2}\dot{\bm{\mu}}\bm{\mu}^{'}
+
\frac{2}{N_0}\Re\{
\dot{\bm{y}}_m(\bm{\gamma}_0)^\mathsf{H}
\bm{y}^{'}_m(\bm{\gamma}_0)
\},
\end{aligned}
\end{equation*}
}
where $\dot{\bm{\mu}}=\Re
\{
\dot{\bm{y}}_m(\bm{\gamma}_0)^\mathsf{H}
(\bm{y}_c-\bm{y}_m(\bm{\gamma}_0))
\}$, and $\bm{\mu}^{'}=\Re
\{
\bm{y}^{'}_m(\bm{\gamma}_0)^\mathsf{H}
(\bm{y}_c-\bm{y}_m(\bm{\gamma}_0))
\}$.
\vspace{-0.1cm}
\subsection{Standard CRB Analysis}
In this subsection, we conduct a standard CRB analysis to derive the theoretical performance when the UE is aware of the presence of MC effects. 
A comparison with the MCRB will highlight the performance degradation when MC effects are neglected by the UE. 
Specifically, we focus on the scenario where the UE is aware of the MC presence but does not know the MC vector $\bm{s}$. 
The corresponding Fisher information matrix denoted by $\bm{J}_2(\bm{\zeta})\in \mathbb{R}^{(2N_m+5)\times (2N_m+5)}$ is defined 
for the state parameter vector $\bm{\zeta}=[\beta_R,\beta_I,\bm{p}_u^T,\bm{s}_m^\mathsf{T}]^\mathsf{T}$, where $\bm{s}_m\in \mathbb{R}^{2N_m}$ with $[\bm{s}_m]_{2i-1}=\Re\{s_i\}$ and $[\bm{s}_m]_{2i}=\Im\{s_i\}$. 
Thus, the PEB is calculated as:
\begin{equation}\label{eq:PEB_aware}
\mathrm{PEB}^{\text{aware}}
=
\mathrm{Tr}\{
[\bm{J}_2^{-1}(\bm{\zeta})]_{3:5,3:5}
\}.
\end{equation}
\vspace{-0.5cm}
\section{Proposed Method}
\vspace{-0.1cm}
To reduce complexity and facilitate optimization, instead of direct localization, we rewrite the UE position as the function of its 2D-AOD $\bm{\psi}_{r,u}=[\theta_{r,u},\phi_{r,u}]^\mathsf{T}$ and distance $d_u$ with respect to the RIS as $\bm{p}_u=\bm{p}_r+d_u\bm{u}(\bm{\psi}_{r,u})$. 
Here, $\theta_{r,u}$ and $\phi_{r,u}$ correspond to the elevation AOD, which measures the angle between $\bm{u}$ and the XY-plane, and the azimuth AOD, which defines the angle between the projection of $\bm{u}$ onto the XY-plane and the X-axis, respectively. 
These angles can be determined as follows:
\begin{equation}\label{eq:theta_phi_vals}
\theta_{r,u}
=
\arccos \left(
\frac{[\bm{p}_{u;r}]_{3}}{\lVert\bm{p}_{u;r}\rVert}
\right),\ 
\phi_{r,u}
=
\text{arctan2} (
[\bm{p}_{u;r}]_{2},[\bm{p}_{u;r}]_{1}
),
\end{equation}
where $\bm{p}_{u;r}=\bm{R}^\mathsf{T}(\bm{p}_u-\bm{p}_{r})$,  
with $\bm{R} \in \mathbb{R}^{3\times 3}$ representing the rotation matrix that defines the orientation of the RIS. 
Moreover,
$
\bm{u}(\bm{\psi})=[\sin\theta \cos\phi,\sin\theta\sin\phi,\cos\theta]^\mathsf{T}
$
is the unit vector pointing in the direction of the 2D-AOD $\bm{\psi} = [\theta, \phi]^\mathsf{T}$. 
Thus, the optimization problem is rewritten as follows:
\begin{equation}\label{eq:optimization-problem-simp-AOD-dist}
(\hat{\bm{\psi}}_{r,u},\hat{d}_u,\hat{\bm{s}})
=
\argmax_{\bm{\psi}_{r,u},d_u,\bm{s}}
\lvert\bm{h}^\mathsf{H}\bm{y}\rvert^2
/
\lVert
\bm{h}
\rVert^2.
\end{equation}
\subsection{Initial Coarse 2D-AOD Estimation}\label{sec:AOD-init}
To obtain an initial estimate of the 2D-AOD $\bm{\psi}_{r,u}$ efficiently, we simplify the problem by assuming an MC-free scenario, i.e., $\bm{s} = \bm{0}_{N_m}$ and $\bm{h}\approx \tilde{\bm{h}}(\bm{\psi}_{r,u},d_u)$. 
Next, we approximate the NF steering vectors in \eqref{eq:NF-str-vec} by using their FF equivalents as \cite{Ozturk2023Phase}:
\begin{equation}
[\bm{a}(
\bm{p}
)]_n
\approx
[\bm{a}_\varphi(
\bm{\psi}_{r,u}
)]_n
=
\text{exp}
\left(
jk
\tilde{\bm{p}}_{r,n}^{\mathsf{T}}
\bm{u}(\bm{\psi}_{r,u})
\right),
\end{equation}
where $\tilde{\bm{p}}_{r,n}$ denotes the local coordinate of the $n$-th RIS element.
Moreover, $\bm{a}_\varphi(.) \in \mathbb{C}^{M_r}$ denotes the RIS FF angle-dependent steering vector.
Under this assumption, the vector $\tilde{\bm{h}}$ can be approximated as:
\begin{equation}\label{eq:h-MC-unaware}
\tilde{\bm{h}}^\mathsf{T}
\approx
\bm{c}^\mathsf{T}(\bm{\psi})
=
(
\bm{a}_r(\bm{\psi})
\odot 
\bm{a}_r(\bm{\psi}_{b,r}^{a})
)^\mathsf{T}
\bm{W}.
\end{equation}
Finally, the 2D-AOD region is discretized into a 2D grid mesh with coarse step sizes $c_\theta$ and $c_\phi$ along the respective dimensions. 
For each grid point $\bm{\psi}_i$, according to the simplified optimization problem in \eqref{eq:optimization-problem-simp}, we compute \eqref{eq:h-MC-unaware} at $\bm{\psi}_i$ and normalize the resulting vector to obtain the vector
$
\frac{\bm{c}(\bm{\psi}_i)}{\lVert \bm{c}(\bm{\psi}_i) \rVert}
$. 
By stacking the normalized vectors, we form the matrix $\bm{C} \in \mathbb{C}^{N_t \times N_c}$. 
To estimate the 2D-AOD for a given observation vector $\bm{y}$, we first calculate the vector $\bm{C}^\mathsf{H} \bm{y}$. 
The corresponding 2D-AOD is then determined by selecting the grid point with the maximum absolute value:
\begin{equation}\label{eq:AOD-init}
\hat{i}
=
\argmax_{i}
\lvert
[\bm{C}]_{:,i}^\mathsf{H} \bm{y}
\rvert
,\ 
\hat{\bm{\psi}}_{r,u}
=
\bm{\psi}_{\hat{i}} \,.
\end{equation}
To improve efficiency, the matrix $\bm{C}$ can be precomputed and reused for initial 2D-AOD estimation at any location.
\subsection{Initial Coarse Distance Estimation}\label{sec:dist-init}
By fixing the 2D-AOD by using the obtained initial estimate in the previous section, and assuming $\bm{s}=\bm{0}_{N_m}$, the distance is estimated via 1D search as follows:
\begin{equation}\label{eq:optimization-problem-simp-d}
\hat{d}_u
=
\argmax_{d_u}
\lvert\tilde{\bm{h}}(\hat{\bm{\psi}}_{r,u},d_u)^\mathsf{H}\bm{y}\rvert^2
/
\lVert
\tilde{\bm{h}}(\hat{\bm{\psi}}_{r,u},d_u)
\rVert^2.
\end{equation}
\subsection{Initial Coarse MC Estimation}\label{sec:MC-init}
We assume a challenging scenario where no channel measurements or prior information are available on MC scattering values. 
In the following we propose a simple closed form approach for initial MC coefficients estimation.
First, we approximate MC affected RIS profiles defined in \eqref{def:Omega-definition} using the first two terms of the Neumann series expansion as \cite{Pinjun2024Impact}:
\begin{align}\label{def:approx-omega}
\bm{\Omega}_{t}^{'}
=
\sum_{n=0}^{\infty}
(\bm{\Omega}_t\bm{S})^n
\bm{\Omega}_t 
& \approx 
\bm{\Omega}_t
+
\bm{\Omega}_t
\bm{S}
\bm{\Omega}_t \notag\\
& =
\bm{\Omega}_t
+
\sum_{i=0}^{N_m-1}
s_i
\bm{\Omega}_t
\bm{A}_i
\bm{\Omega}_t
\end{align}

Thus, $\bm{h}$ in \eqref{eq:ch-element} can be approximated as:
\begin{equation}\label{eq:ch-element-approx}
\bm{h}
=
\tilde{\bm{h}}
+
\bm{G}
\bm{s},
\end{equation}
where $\bm{G}\in\mathbb{C}^{N_t\times N_m}$ with the $(t,i)$-th element 
$$[\bm{G}]_{t,i}=\bm{a}(\bm{p}_u)^\mathsf{T}
\bm{\Omega}_t\bm{A}_i\bm{\Omega}_t
\bm{a}(\bm{p}_b).$$
Hence, the MC vector $\bm{s}$ in \eqref{eq:optimization-problem} can be estimated via least squares as:
\begin{equation}\label{eq:MC-init}
\hat{\bm{s}}
=
\bm{G}^\dagger
(\bm{y}/\beta-\tilde{\bm{h}}),
\end{equation}
where $\tilde{\bm{h}}$ can be estimated using the initially estimated channel parameters discussed in previous subsections. 
The channel gain $\beta$ can be estimated by substituting $\tilde{\bm{h}}$ in \eqref{eq:beta-opt}.
\vspace{-0.1cm}
\subsection{Joint Position and MC Parameter Refinement}
In this subsection, we adopt an alternating optimization approach for the joint refinement of the 2D-AOD, distance, and MC vector. 
Starting with the initial estimates derived in the previous subsections, we iteratively optimize the objective function in \eqref{eq:optimization-problem-simp-AOD-dist} by alternately updating the 2D-AOD, distance, and MC vector until convergence. 
Convergence is achieved when the maximum update of all parameters falls below a predefined threshold $\epsilon$.
The overall proposed JLMC algorithm is summarized in Algorithm~\ref{alg:proposed-alg}. 
It is noteworthy that for steps 6, 7, and 8 of the refinement stage, the off-the-shelf quasi-Newton approach is employed for optimization. 
Upon convergence of the algorithm, the 3D position of the UE in the local coordinate system is determined as
\begin{equation}\label{eq:pos-est-local}
\tilde{\bm{p}}_{u}
=
\hat{d}_u
\hat{\bm{u}}_n,
\end{equation}
Here, $\hat{\bm{u}}_n$ represents the unit vector corresponding to the estimated 2D-AOD $\hat{\bm{\psi}}_{r,u}$, and $\tilde{\bm{p}}_{u}$ denotes the estimated location in the RIS's local coordinate system. 
The global coordinate position can then be readily computed as
\begin{equation}\label{eq:pos-est-global}
\hat{\bm{p}}_{u}
=
\bm{R}
\tilde{\bm{p}}_{u}
+
\bm{p}_r.
\end{equation}

\begin{algorithm}
\caption{Proposed JLMC Algorithm}\label{alg:proposed-alg}
\begin{algorithmic}[1]
\State 
\multiline{%
\textbf{Inputs}:
Received signal $\bm{y}$ measured at unknown position $\bm{p}_u$ based on \eqref{eq:Y}, and convergence threshold $\epsilon$.
}
\State 
\multiline{%
\textbf{Output}: 
Estimated 3D position $\hat{\bm{p}}_u$ and MC vector $\hat{\bm{s}}$.
}
\State 
\multiline{%
\textbf{Initialization Stage:} 
Find an initial estimate of the 2D-AOD $\hat{\bm{\psi}}_{r,u}$, distance $\hat{d_u}$, and the MC vector $\hat{\bm{s}}$ according to \eqref{eq:AOD-init}, \eqref{eq:optimization-problem-simp-d}, and \eqref{eq:MC-init}, respectively.
}
\State
\multiline{%
\textbf{Refinement Stage:} 
Refine the estimated parameters via alternating optimization approach as follows:
}
\While{maximum parameter update is greater than $\epsilon$}
\State 
\multiline{%
Update the 2D-AOD $\hat{\bm{\psi}}_{r,u}$ by maximizing \eqref{eq:optimization-problem-simp-AOD-dist} by keeping the other variables fixed.
}
\State 
\multiline{%
Update the distance $\hat{d}_u$ by maximizing \eqref{eq:optimization-problem-simp-AOD-dist} by keeping the other variables fixed.
}
\State 
\multiline{%
Update the MC vector $\bm{s}$ by maximizing \eqref{eq:optimization-problem-simp-AOD-dist} by keeping the other variables fixed.
}
\EndWhile
\State 
\multiline{%
Estimate the UE position $\bm{p}_u$ using \eqref{eq:pos-est-global}.
}
\end{algorithmic}
\end{algorithm}

\section{Simulation Results}
\vspace{-0.1cm}
In this section, we perform simulations to validate the proposed JLMC algorithm in RIS-assisted ISAC systems. 
The default parameters are shown in Table~\ref{tab:sys-params}. 
The chosen RIS size $M_1 = M_2 = 48$ results in a NF region spanning from $1.187\,\mathrm{m}$ to $22.074\,\mathrm{m}$. 
Furthermore, the elements of the RIS phase profiles, $\{\bm{\Omega}_t\}_{t=1}^{N_t}$, are randomly selected from the range $[-\pi, \pi]$.
In all simulations, the number of MC coefficients is set to $N_m = 3$. 
The default unit-norm MC vector, $\bm{s}$, is specified in Table~\ref{tab:sys-params}. 
For cases where $\lVert \bm{s} \rVert = q$, the default vector from Table~\ref{tab:sys-params} is scaled by $q$ to achieve the desired norm. The received signal is generated using \eqref{eq:Y}.
\vspace{-0.2cm}
\begin{table}[ht]
\caption{System parameters}
\centering
\fontsize{12}{10}\selectfont 
\resizebox{\columnwidth}{!}{
\begin{tabular}{ |l|l|l|  }
\hline
\textbf{System Parameters}
&
\textbf{Symbol}
&
\textbf{Default Value}
\\
\hline
Number of pilot symbols & $N_t$ & $15$
\\
Carrier frequency & $f_c$ & $30\,\mathrm{GHz}$
\\
Bandwidth & $W$ & $1\,\mathrm{MHz}$
\\
Noise PSD & $N_0$ & $-173.855\,\mathrm{dBm}$
\\
Transmit antenna gain & $G_t$ & $1$
\\
Receive antenna gain & $G_r$ & $1$
\\
Light speed & $c$ & $3\times 10^8$
\\
MC vector & 
$\bm{s}$ &  
$
\begin{aligned}[t]
& [-0.681+0.458j, -0.506+0.0492j, \\
& 0.244+0.0928j]^\mathsf{T}
\end{aligned}
$
\\
BS position & $\bm{p}_b$ & $[0, 0, 2.5]^\mathsf{T}\,\mathrm{[m]}$
\\
RIS position & $\bm{p}_{r}$ & $[0, 5, 2]^\mathsf{T}\,\mathrm{[m]}$
\\
UE position & $\bm{p}_{u}$ & $[7, 3, 1.5]^\mathsf{T}\,\mathrm{[m]}$
\\
RIS number of elements & $M_r$ & $48\times 48=2304$
\\
Algorithm stop threshold & $\epsilon$ & $10^{-15}$
\\
\hline
\end{tabular}
}
\label{tab:sys-params}
\end{table}

\subsection{Performance of the Proposed JLMC}\label{sec:JLMC_performance}
This subsection examines the performance of the proposed JLMC algorithm by analyzing the root mean squared error (RMSE) across various power levels and MC severity conditions. 
To provide a benchmark, the values $\mathrm{PEB}^{\text{unaware}}$ and $\mathrm{PEB}^{\text{aware}}$, as defined in \eqref{eq:PEB_unaware} and \eqref{eq:PEB_aware}, are used for comparison. 
The JLMC algorithm, described in Algorithm~\ref{alg:proposed-alg}, jointly estimates both the MC parameters and the location. 
Simulations are conducted for $\lVert \bm{s} \rVert \in \{0, 0.01, 0.05\}$ and power levels $P \in \{-15, -10, -5, 0, 5, 10\}\,\mathrm{dBm}$, with $N_{\text{gen}} = 500$ data samples generated for each scenario. 

\begin{figure}
\centering
\includegraphics[width=\columnwidth]{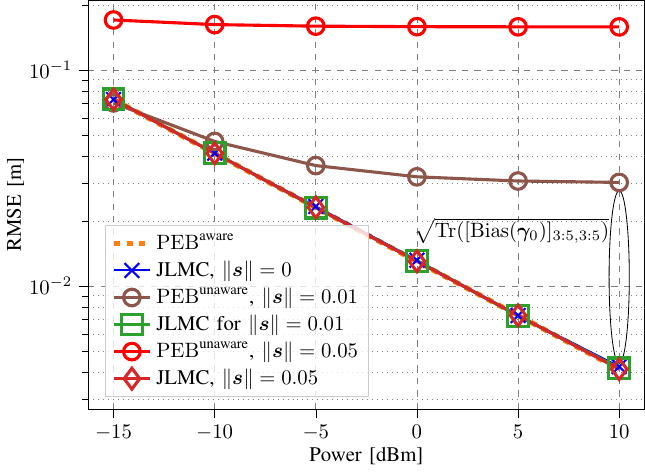}
\caption{
Comparison of the proposed JLMC algorithm and the lower bounds.
}
\label{fig:RMSE_loc}
\end{figure}

The localization RMSE results are illustrated in Fig.~\ref{fig:RMSE_loc}.
As shown in the figure, higher values of $\lVert \bm{s} \rVert$, corresponding to more pronounced MC effects, lead to significant deviations of conventional MC-unaware methods from the true CRB. 
By contrast, the proposed MC-aware JLMC algorithm exhibits strong alignment with the CRB, demonstrating its robustness. 
It is worth noting that since the $\mathrm{PEB}^{\text{aware}}$ curves for different $\lVert \bm{s} \rVert$ are nearly identical, only the curve for $\lVert \bm{s} \rVert = 0$ is plotted for simplicity. 
This is due to the fact that the MC scattering components in the vector $\bm{s}$ are nuisance and do not contribute meaningful information to the localization process. 
This observation will be further illustrated in the subsequent subsections.
Furthermore, the bias term, defined in \eqref{eq:LB_def}, is shown for $\lVert \bm{s} \rVert = 0.01$.

\begin{figure}[!t]
\centering
\includegraphics[width=\columnwidth]{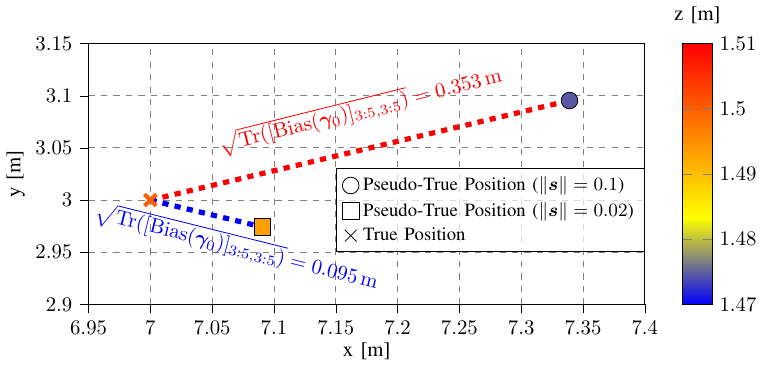}
\caption{
Pseudo-true UE positions for $\lVert \bm{s}\rVert\in\{0.02,0.1\}$.
The colorbar shows the z coordinate values.
}
\label{fig:UE_locs}
\end{figure}

Fig.~\ref{fig:UE_locs} illustrates the pseudo-true UE positions for two MC vector norms, $\lVert\bm{s}\rVert=0.02$ and $\lVert\bm{s}\rVert=0.1$, along with the corresponding bias terms. 
It is noteworthy that classical MC-unaware methods converge to these positions as the transmit power increases, with their RMSE values approaching the associated bias terms, highlighting the significant performance degradation of MC-unaware approaches.

\vspace{-0.1cm}
\subsection{Impact of $\lVert \bm{s}\rVert$ on JLMC and Benchmarks}
\begin{figure}[!t]
\centering
\includegraphics[width=\columnwidth]{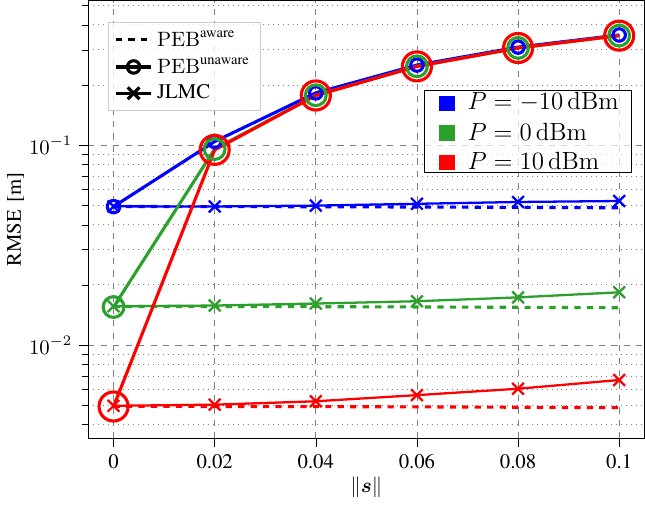}
\caption{
Localization performance versus MC severeness.
}

\label{fig:PEB_mc_norm}
\end{figure}

In this experiment, we evaluate the effect of $\lVert \bm{s} \rVert$, which quantifies the severity of MC, on system performance. 
Specifically, RMSE results and benchmarks are obtained for $0 \leq \lVert \bm{s} \rVert \leq 0.1$, as shown in Fig.~\ref{fig:PEB_mc_norm}. 
The results indicate that the CRB values are approximately identical with respect to $\lVert \bm{s} \rVert$, validating our claim in Section~\ref{sec:JLMC_performance}. 
In contrast, the $\mathrm{PEB}^{\text{unaware}}$ values grow significantly with higher $\lVert \bm{s} \rVert$, demonstrating that conventional MC-unaware methods become increasingly suboptimal as MC effects are neglected. 

Additionally, consistent with observations in the previous subsection, $\mathrm{PEB}^{\text{unaware}}$ remains invariant across the power levels considered. 
This behavior arises because, beyond a certain power, $\mathrm{PEB}^{\text{unaware}}$ converges to the bias term defined in \eqref{eq:LB_def} (the MCRB term converges to zero). 
On the other hand, the proposed JLMC algorithm closely follows the CRB curves, effectively mitigating MC-related degradation. 
However, it is observed that the performance of JLMC deteriorates slightly with larger $\lVert \bm{s} \rVert$, leading to minor deviations from the CRB. 
This degradation can be attributed to reduced accuracy in the initialization steps of JLMC, detailed in Sections~\ref{sec:AOD-init}, \ref{sec:dist-init}, and \ref{sec:MC-init}. 
Since the quasi-Newton optimization approach is highly sensitive to initialization precision, the performance impact becomes noticeable for higher $\lVert \bm{s} \rVert$. 
Nevertheless, the proposed JLMC algorithm consistently outperforms $\mathrm{PEB}^{\text{unaware}}$ and remains close to the CRB curves.

\vspace{-0.1cm}
\section{Conclusion}\label{sec:conclusion}
This paper addressed the problem of 3D localization for a single-antenna UE in a NF RIS-assisted ISAC system, considering the impact of MC among RIS elements. 
We proposed a two-stage framework for joint UE localization and MC parameter estimation. 
In the first stage, initial estimates of the 2D-AOD, UE-RIS distance, and MC parameters were obtained, while in the second stage, these estimates were refined using an alternating optimization approach. 
The performance of the proposed method was analyzed through MCRB and standard CRB derivation via FIM analysis, which highlighted the effectiveness of the proposed approach and the significant degradation of classical MC-unaware techniques. 
Simulation results validated the accuracy and efficiency of the proposed algorithm. 
Future work will explore scenarios involving multiple simultaneous hardware impairments to further enhance the system's robustness in practical environments.
\vspace{-0.05cm}
\section*{Acknowledgment}
This work is supported by the SNS JU project 6G-DISAC under the EU's Horizon Europe research and innovation program under Grant Agreement No 101139130, and by the Swedish Research Council (VR) through the project 6G-PERCEF under Grant 2024-04390.
\vspace{-0.05cm}
\bibliographystyle{IEEEtran}
 \bibliography{Bib}

\end{document}